\documentclass{article} 
\usepackage{iclr2024_conference,times}

\usepackage[hidelinks]{hyperref}
\usepackage{subfigure}
\usepackage{fancyvrb}
\usepackage{amsmath, amsfonts, amsthm, amssymb}
\usepackage{stmaryrd}
\usepackage{verbatim}
\usepackage{graphicx}
\usepackage{setspace}
\usepackage[ruled, vlined]{algorithm2e}
\graphicspath{{Figures/}}

\usepackage{listings}
\usepackage{xcolor}
\definecolor{codegreen}{rgb}{0,0.6,0}
\definecolor{codegray}{rgb}{0.5,0.5,0.5}
\definecolor{codepurple}{rgb}{0.58,0,0.82}
\definecolor{backcolour}{rgb}{0.95,0.95,0.92}

\lstdefinestyle{mystyle}{
    backgroundcolor=\color{backcolour},   
    commentstyle=\color{codegreen},
    keywordstyle=\color{magenta},
    numberstyle=\tiny\color{codegray},
    stringstyle=\color{codepurple},
    basicstyle=\ttfamily\footnotesize,
    breakatwhitespace=false,         
    breaklines=true,                 
    captionpos=b,                    
    keepspaces=true,                 
    numbers=left,                    
    numbersep=5pt,                  
    showspaces=false,                
    showstringspaces=false,
    showtabs=false,                  
    tabsize=2
}
\theoremstyle{definition}

\numberwithin{equation}{section}

\title{AI-driven Alternative Medicine:\\A Novel Approach to Drug Discovery\\and Repurposing}

\author{~\\\textbf{Oleksandr BILOKON} (Founder, Managing Director, Thalesians Marine Ltd) \\ \textbf{Nataliya BILOKON} (Adviser, Thalesians Marine Ltd) \\ \textbf{Paul Alexander BILOKON} (Chief Scientific Adviser, Thalesians Marine Ltd)}

\usepackage{tikz}
\usetikzlibrary{positioning}

\iclrfinalcopy 
\begin{document}

\maketitle

\begin{abstract}
AIAltMed is a cutting-edge platform designed for drug discovery and repurposing. It utilizes Tanimoto similarity to identify structurally similar non-medicinal compounds to known medicinal ones. This preprint introduces AIAltMed, discusses the concept of `AI-driven alternative medicine,' evaluates Tanimoto similarity's advantages and limitations, and details the system's architecture. Furthermore, it explores the benefits of extending the system to include PubChem and outlines a corresponding implementation strategy.
\end{abstract}

\section{Introduction}
Drug discovery and repurposing are critical areas of pharmaceutical research. Traditional methods are often time-consuming and costly. AIAltMed (\url{http://aialtmed.com/}) presents a novel solution by utilizing AI and Tanimoto similarity (see~\cite{tanimoto1958elementary}), also known as the Jaccard index (see~\cite{jaccard1901distribution}), to accelerate these processes. This platform not only aids in identifying potential new uses for existing drugs but also explores non-medicinal compounds with therapeutic potential, thus pioneering the field of AI-driven alternative medicine.

\section{AI-driven Alternative Medicine}
AI-driven alternative medicine is a novel approach that leverages artificial intelligence to identify non-medicinal compounds with structural similarities to medicinal compounds. By using Tanimoto similarity, AIAltMed can find compounds in databases like DrugBank (see~\cite{Knox2023}) and FooDB (see~\cite{FooDB}) that may exhibit therapeutic effects. This method opens new avenues for discovering alternative treatments and broadening the scope of drug repurposing.

This approach has been inspired by~\cite{MacMahon2023}, where an \textit{in silico} drug repurposing pipeline was developed to identify drugs with the potential to inhibit SARS-CoV-2 replication, based on structural similarity to drugs already in clinical trials for COVID-19, leading to the identification of two candidate drugs for repurposing with the potential to inhibit SARS-CoV-2 replication (triamcinolone and gallopamil). The subsequent analysis proposed \textit{ibid.} based on affected pathways is a natural next step to the similarity search.

This approach has similarities to that discussed in~\cite{Szilagyi2021}. A review of similarity-based approaches to molecular structure can be found in~\cite{Bero2017}. 

\section{Tanimoto Similarity and Similarity-based Molecule Search}

Tanimoto similarity is a measure of the structural similarity between two molecules, calculated as the ratio of the intersection over the union of their molecular fingerprints.

\subsection{Advantages}
\begin{itemize}
    \item Efficiently identifies structurally similar compounds.
    \item Facilitates rapid screening of large compound libraries.
    \item Helps in predicting biological activity based on structural similarity.
\end{itemize}

\subsection{Disadvantages}
\begin{itemize}
    \item May overlook compounds with different structures but similar biological activity.
    \item Dependent on the quality and completeness of the molecular fingerprint database.
    \item Potential for high false positive rates if not carefully validated.
\end{itemize}

Further merits and demerits of Tanimoto-similarity are discussed in~\cite{Bajusz2015}.

\section{System Architecture}
The AIAltMed system is built using Django (see~\cite{django}), a high-level Python (see~\cite{python}) web framework, and caches the similarity table in memory for efficient retrieval and processing.

\begin{figure}[h]
    \centering
    \includegraphics[width=0.8\textwidth]{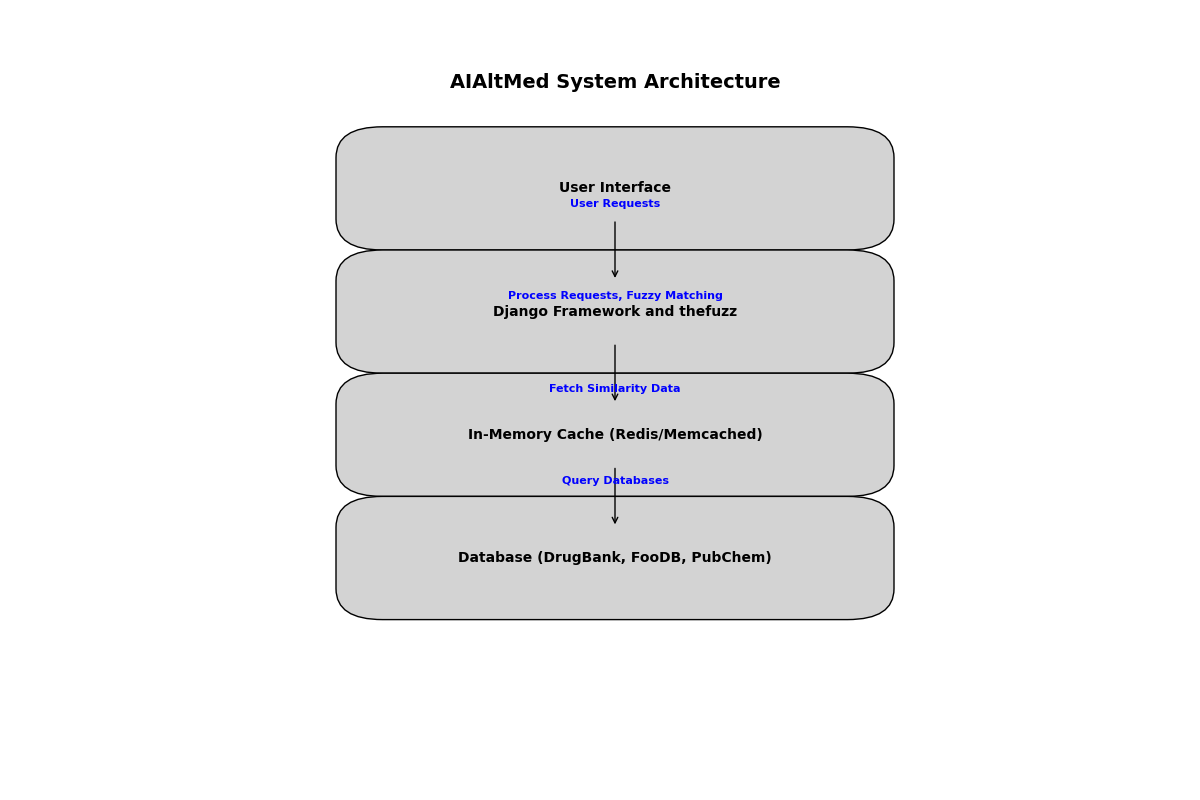}
    \caption{System Architecture of AIAltMed}
    \label{fig:architecture}
\end{figure}

\subsection{Django Framework}

\sloppy Django provides a robust and scalable framework for developing the AIAltMed platform, offering advantages such as rapid development, security features, and a comprehensive set of tools.

\fussy Django is a high-level Python web framework that is well-suited for the development of the AIAltMed application due to its rapid development capabilities and robust feature set. One of the primary advantages of Django is its "batteries-included" philosophy, which provides a comprehensive set of tools and libraries out-of-the-box. This allows developers to focus on building the core functionality of AIAltMed without needing to spend extensive time on setting up and configuring ancillary components. Django's built-in admin interface is particularly useful for managing the application's data and user authentication, streamlining the process of creating and maintaining the backend. Additionally, Django's ORM (Object-Relational Mapping) system simplifies database interactions, enabling efficient handling of complex queries and ensuring data integrity.

Another significant advantage of Django is its emphasis on security. Django includes numerous built-in security features, such as protection against SQL injection, cross-site scripting (XSS), cross-site request forgery (CSRF), and clickjacking. These features are essential for ensuring that the AIAltMed application can securely handle sensitive data, including proprietary chemical information and user credentials. Django's scalability is also noteworthy; it supports the development of applications ranging from simple websites to complex, data-intensive platforms like AIAltMed. The framework's ability to handle high traffic and large datasets ensures that AIAltMed can grow and scale as its user base and data volume expand. Furthermore, Django's extensive documentation and active developer community provide valuable resources and support, facilitating continuous improvement and troubleshooting.

\subsection{Fuzzy Search Implementation}

The Django/Python implementation of the AIAltMed system leverages fuzzy search functionality provided by the \texttt{thefuzz} library, a popular Python package available on GitHub. \texttt{thefuzz} (formerly known as \texttt{fuzzywuzzy}) is an effective tool for approximate string matching. It is widely used in applications where exact string matches are not always possible or practical. By employing the Levenshtein distance algorithm, \texttt{thefuzz} can calculate the similarity between two strings, allowing the AIAltMed system to find and rank potential matches based on their resemblance. This capability is crucial for searching large databases where slight variations in data entry or terminology might otherwise lead to missed results.

Integrating \texttt{thefuzz} into the AIAltMed system enhances its ability to accurately retrieve relevant compounds, even when the search terms are not an exact match to the database entries. This is particularly beneficial in the context of drug discovery and repurposing, where the chemical names and descriptions might have minor inconsistencies. The use of fuzzy search ensures that researchers can identify all pertinent compounds without being hindered by minor discrepancies in naming conventions. Overall, \texttt{thefuzz} significantly improves the robustness and reliability of the AIAltMed system's search capabilities, making it a more effective tool for identifying potential therapeutic compounds.

\subsection{In-memory Caching}

In-memory caching (see~\cite{metzger2011in}) the similarity table in memory allows for fast access and processing, significantly improving the performance of similarity searches.

In-memory caching is a powerful technique that significantly enhances the performance and efficiency of web applications by storing frequently accessed data in memory rather than on disk. This approach reduces the latency associated with retrieving data from traditional storage systems, enabling faster response times and improving the overall user experience. For the AIAltMed application, in-memory caching plays a crucial role in managing the similarity table used for compound searches. By keeping this data readily available in memory, the system can quickly perform complex similarity calculations and retrieve relevant results without the overhead of repeated database queries. This is particularly important given the potentially large size of the similarity table and the need for rapid, real-time access during user interactions.

The use of in-memory caching also contributes to the scalability and reliability of the AIAltMed system. As the number of users and the volume of data grow, the demand on the backend infrastructure increases. In-memory caching helps to mitigate this by offloading a significant portion of the read operations from the database to the cache, thereby reducing the load on the database and improving its overall performance. Additionally, in-memory caching solutions, such as Redis or Memcached, offer features like data persistence, replication, and failover, which enhance the system's fault tolerance and data durability. These features ensure that the AIAltMed application remains responsive and available even under high load conditions or in the event of hardware failures. By leveraging in-memory caching, AIAltMed can maintain high performance and reliability, supporting its mission to provide efficient and effective drug discovery and repurposing solutions.

\section{Extending to PubChem}

Extending AIAltMed to include all of PubChem (see~\cite{Kim2022}) would vastly increase the potential for discovering novel therapeutic compounds. PubChem is one of the largest repositories of chemical information, and integrating it into AIAltMed could enhance the system's ability to identify relevant compounds.

\subsection{Advantages}
\begin{itemize}
    \item Access to a more extensive and diverse chemical library.
    \item Increased likelihood of identifying novel compounds with therapeutic potential.
    \item Enhanced validation and cross-referencing capabilities.
\end{itemize}

\subsection{Implementation Strategy}
\begin{itemize}
    \item Update the similarity search algorithm to handle larger datasets efficiently.
    \item Implement scalable storage solutions to manage the increased data volume.
    \item Optimize the in-memory caching mechanism to ensure performance is maintained.
    \item Collaborate with PubChem to ensure data integration and consistency.
\end{itemize}

\section{Scaling Up}

Scaling up AIAltMed, given its current architecture, involves several strategic enhancements across various components of the system. Firstly, to handle increased user demand and data volume, the backend infrastructure must be robust and scalable. This can be achieved by leveraging cloud computing platforms such as AWS, Azure, or Google Cloud, which offer scalable compute and storage resources. Using these platforms, AIAltMed can dynamically allocate resources based on demand, ensuring that the application remains responsive even during peak usage. Additionally, implementing containerization with Docker and orchestration with Kubernetes can streamline the deployment process, allowing the system to scale horizontally by adding more instances of the Django application as needed.

Incorporating advanced database management techniques is another critical step in scaling AIAltMed. The current use of in-memory caching with Redis or Memcached significantly enhances performance by reducing latency in data retrieval. To further improve scalability, the database architecture could adopt a distributed database system such as Apache Cassandra or Google Bigtable, which can handle large-scale data across multiple nodes, ensuring high availability and fault tolerance. Furthermore, optimizing the existing database queries and indexing strategies can minimize the load on the database and speed up data access times, making the system more efficient.

Finally, integrating machine learning models like Graph Neural Networks (GNNs) can enhance the system’s capability to process and analyze large datasets effectively. GNNs can provide deeper insights into molecular interactions and predict the therapeutic potential of compounds more accurately. To support the computational requirements of these models, AIAltMed can utilize GPU-accelerated computing or specialized hardware like TPUs (Tensor Processing Units) available on cloud platforms. Additionally, implementing a microservices architecture can decouple different functionalities of the application, allowing independent scaling of each service. For instance, the fuzzy search service using thefuzz can be scaled independently from the main Django application, ensuring that each component operates efficiently under increased load. These strategies collectively enable AIAltMed to scale effectively, accommodating growth in user base, data volume, and computational complexity.

\section{Conclusion}

AIAltMed marks a substantial leap in drug discovery and repurposing. Utilizing AI and Tanimoto similarity, it efficiently identifies both medicinal and non-medicinal compounds with therapeutic potential. Expanding the system to integrate PubChem will significantly enhance its capabilities, offering an invaluable resource for researchers and clinicians in developing new treatments. This innovative approach underscores the potential of AI-driven alternative medicine to revolutionize pharmaceutical research and personalized healthcare.

Future development of AIAltMed could explore the integration of graph neural networks (GNNs)~\cite{zhou2020graph} to enhance the system's ability to identify and predict the therapeutic potential of compounds. GNNs, which are designed to process data structured as graphs, can be particularly effective for analyzing molecular structures, which are naturally represented as graphs with atoms as nodes and chemical bonds as edges. By leveraging GNNs, AIAltMed can more accurately model the complex relationships and interactions within molecular structures, potentially uncovering new insights into how different compounds might interact with biological targets. This advanced modeling could improve the system's ability to predict biological activity and identify novel drug candidates, thus extending the platform's utility in drug discovery and repurposing.

Additionally, AIAltMed could benefit from expanding its dataset to include more diverse and comprehensive chemical libraries. Incorporating data from additional databases beyond DrugBank, FooDB, and PubChem, such as ChEMBL~\cite{gaulton2012chembl} or ZINC~\cite{irwin2012zinc}, would provide a richer set of molecular structures for analysis. Coupled with enhanced computational techniques like GNNs, this expanded dataset could enable the system to identify more subtle patterns and relationships within the data, leading to more robust predictions and discoveries. Furthermore, AIAltMed could explore the use of federated learning to collaborate with other research institutions and pharmaceutical companies. This approach would allow the platform to leverage vast amounts of distributed data while maintaining data privacy and security, fostering innovation and accelerating the discovery of new therapeutic compounds.

\section{Acknowledgements}

This work was performed at Thalesians Marine Ltd, a private limited company registered in England and Wales with company number 12147626 with registered office address 3rd Floor, 120 Baker Street, London, England, W1U 6TU and trading address Level39, One Canada Square, Canary Wharf, London, England, E14 5AB.

Level39, located in the heart of London's Canary Wharf, is one of the world's most connected technology hubs, fostering innovation and growth for startups and scaleups in the fields of finance, cybersecurity, retail, and smart cities. As a member of the Level39 community, Thalesians Marine Ltd benefits from access to a rich ecosystem of industry experts, investors, and mentors who provide invaluable support and guidance.

The authors would like to thank Thalesians Marine Ltd and Level39 for creating a working environment conducive to innovative research and development work.

\section{Disclaimer}

The information provided in this work is for general informational purposes only. It is not intended as a substitute for professional medical advice, diagnosis, or treatment. Always seek the advice of your physician or other qualified healthcare provider with any questions you may have regarding a medical condition. Never disregard professional medical advice or delay in seeking it because of something you have read in this work. Reliance on any information provided in this work is solely at your own risk. The authors of this work do not assume any responsibility or liability for any injury, damage, or other adverse consequences resulting from the use or misuse of the information provided herein. Furthermore, the information in this work may not be applicable to your particular medical condition or health concern. It is important to consult with a qualified healthcare professional before making any decisions about your health or medical treatment. The inclusion of any links to preprints, research papers, and third-party websites does not imply endorsement or recommendation by the authors. We are not responsible for the content or accuracy of any preprints, research papers, and third-party websites referenced from this work. Please consult with your healthcare provider if you have any questions or concerns about your health or medical condition.


\begin{thebibliography}{15}
\providecommand{\natexlab}[1]{#1}
\providecommand{\url}[1]{\texttt{#1}}
\expandafter\ifx\csname urlstyle\endcsname\relax
  \providecommand{\doi}[1]{doi: #1}\else
  \providecommand{\doi}{doi: \begingroup \urlstyle{rm}\Url}\fi

\bibitem[Bajusz et~al.(2015)Bajusz, Rácz, and Héberger]{Bajusz2015}
Dávid Bajusz, Anita Rácz, and Károly Héberger.
\newblock Why is tanimoto index an appropriate choice for fingerprint-based similarity calculations?
\newblock \emph{Journal of Cheminformatics}, 7\penalty0 (1), May 2015.
\newblock ISSN 1758-2946.
\newblock \doi{10.1186/s13321-015-0069-3}.

\bibitem[Bero et~al.(2017)Bero, Muda, Choo, Muda, and Pratama]{Bero2017}
S~A Bero, A~K Muda, Y~H Choo, N~A Muda, and S~F Pratama.
\newblock Similarity measure for molecular structure: A brief review.
\newblock \emph{Journal of Physics: Conference Series}, 892:\penalty0 012015, September 2017.
\newblock ISSN 1742-6596.
\newblock \doi{10.1088/1742-6596/892/1/012015}.

\bibitem[Foundation(2005)]{django}
Django~Software Foundation.
\newblock Django: The web framework for perfectionists with deadlines.
\newblock \url{https://www.djangoproject.com/}, 2005.
\newblock Accessed: 2024-07-02.

\bibitem[Foundation(1991)]{python}
Python~Software Foundation.
\newblock Python programming language.
\newblock \url{https://www.python.org/}, 1991.
\newblock Accessed: 2024-07-02.

\bibitem[Gaulton et~al.(2012)Gaulton, Bellis, Bento, Chambers, Davies, Hersey, Light, McGlinchey, Michalovich, Al-Lazikani, and Overington]{gaulton2012chembl}
Anna Gaulton, Louisa~J Bellis, A~Patricia Bento, Jon Chambers, Mark Davies, Anne Hersey, Yvonne Light, Sean McGlinchey, David Michalovich, Bissan Al-Lazikani, and John~P Overington.
\newblock Chembl: a large-scale bioactivity database for drug discovery.
\newblock \emph{Nucleic acids research}, 40\penalty0 (D1):\penalty0 D1100--D1107, 2012.

\bibitem[Irwin et~al.(2012)Irwin, Sterling, Mysinger, Bolstad, and Coleman]{irwin2012zinc}
John~J Irwin, Teague Sterling, Michael~M Mysinger, E~Lilly Bolstad, and Tara~P Coleman.
\newblock Zinc 15--ligand discovery for everyone.
\newblock \emph{Journal of chemical information and modeling}, 52\penalty0 (7):\penalty0 1757--1768, 2012.

\bibitem[Jaccard(1901)]{jaccard1901distribution}
Paul Jaccard.
\newblock Étude comparative de la distribution florale dans une portion des alpes et des jura.
\newblock \emph{Bulletin de la Société Vaudoise des Sciences Naturelles}, 37:\penalty0 547--579, 1901.

\bibitem[Kim et~al.(2022)Kim, Chen, Cheng, Gindulyte, He, He, Li, Shoemaker, Thiessen, Yu, Zaslavsky, Zhang, and Bolton]{Kim2022}
Sunghwan Kim, Jie Chen, Tiejun Cheng, Asta Gindulyte, Jia He, Siqian He, Qingliang Li, Benjamin~A Shoemaker, Paul~A Thiessen, Bo~Yu, Leonid Zaslavsky, Jian Zhang, and Evan~E Bolton.
\newblock Pubchem 2023 update.
\newblock \emph{Nucleic Acids Research}, 51\penalty0 (D1):\penalty0 D1373--D1380, October 2022.
\newblock ISSN 1362-4962.
\newblock \doi{10.1093/nar/gkac956}.

\bibitem[Knox et~al.(2023)Knox, Wilson, Klinger, Franklin, Oler, Wilson, Pon, Cox, Chin, Strawbridge, Garcia-Patino, Kruger, Sivakumaran, Sanford, Doshi, Khetarpal, Fatokun, Doucet, Zubkowski, Rayat, Jackson, Harford, Anjum, Zakir, Wang, Tian, Lee, Liigand, Peters, Wang, Nguyen, So, Sharp, da Silva, Gabriel, Scantlebury, Jasinski, Ackerman, Jewison, Sajed, Gautam, and Wishart]{Knox2023}
Craig Knox, Mike Wilson, Christen M Klinger, Mark Franklin, Eponine Oler, Alex Wilson, Allison Pon, Jordan Cox, Na~Eun (Lucy) Chin, Seth A Strawbridge, Marysol Garcia-Patino, Ray Kruger, Aadhavya Sivakumaran, Selena Sanford, Rahil Doshi, Nitya Khetarpal, Omolola Fatokun, Daphnee Doucet, Ashley Zubkowski, Dorsa Yahya Rayat, Hayley Jackson, Karxena Harford, Afia Anjum, Mahi Zakir, Fei Wang, Siyang Tian, Brian Lee, Jaanus Liigand, Harrison Peters, Ruo~Qi (Rachel) Wang, Tue Nguyen, Denise So, Matthew Sharp, Rodolfo da Silva, Cyrella Gabriel, Joshua Scantlebury, Marissa Jasinski, David Ackerman, Timothy Jewison, Tanvir Sajed, Vasuk Gautam, and David S Wishart.
\newblock Drugbank 6.0: the drugbank knowledgebase for 2024.
\newblock \emph{Nucleic Acids Research}, 52\penalty0 (D1):\penalty0 D1265--D1275, November 2023.
\newblock ISSN 1362-4962.
\newblock \doi{10.1093/nar/gkad976}.

\bibitem[MacMahon et~al.(2023)MacMahon, Hwang, Yim, MacMahon, Abraham, Barton, Tharmakulasingam, Bilokon, Gaddi, and Han]{MacMahon2023}
Méabh MacMahon, Woochang Hwang, Soorin Yim, Eoghan MacMahon, Alexandre Abraham, Justin Barton, Mukunthan Tharmakulasingam, Paul Bilokon, Vasanthi~Priyadarshini Gaddi, and Namshik Han.
\newblock An in silico drug repurposing pipeline to identify drugs with the potential to inhibit sars-cov-2 replication.
\newblock \emph{Informatics in Medicine Unlocked}, 43:\penalty0 101387, 2023.
\newblock ISSN 2352-9148.
\newblock \doi{10.1016/j.imu.2023.101387}.

\bibitem[Metzger \& Leymann(2011)Metzger and Leymann]{metzger2011in}
Andreas Metzger and Frank Leymann.
\newblock \emph{In-Memory Data Management: Technology and Applications}.
\newblock Springer, 2011.

\bibitem[Szilágyi et~al.(2021)Szilágyi, Flachner, Hajdú, Szaszkó, Dobi, Lőrincz, Cseh, and Dormán]{Szilagyi2021}
Katalin Szilágyi, Beáta Flachner, István Hajdú, Mária Szaszkó, Krisztina Dobi, Zsolt Lőrincz, Sándor Cseh, and György Dormán.
\newblock Rapid identification of potential drug candidates from multi-million compounds’ repositories. combination of 2d similarity search with 3d ligand/structure based methods and in vitro screening.
\newblock \emph{Molecules}, 26\penalty0 (18):\penalty0 5593, September 2021.
\newblock ISSN 1420-3049.
\newblock \doi{10.3390/molecules26185593}.

\bibitem[Tanimoto(1958)]{tanimoto1958elementary}
T.T. Tanimoto.
\newblock An elementary mathematical theory of classification and prediction.
\newblock \emph{International Business Machines Corporation, New York}, pp.\  1--16, 1958.

\bibitem[Wishart et~al.(2018)Wishart, Guo, Oler, Wang, Anjum, Peters, Liang, Vázquez-Fresno, Sajed, Johnson, Karu, Sayeeda, Lo, Gautam, Torres-Calzada, Hameed, LeVatte, Forsythe, and Salek]{FooDB}
David~S. Wishart, An~Chi Guo, Eponine Oler, Fan Wang, Amina Anjum, Justin Peters, Kai Liang, Rosa Vázquez-Fresno, Tanvir Sajed, Daniel Johnson, Naila Karu, Zeinab Sayeeda, Emmeline Lo, Vineet Gautam, Cristhian Torres-Calzada, Imran Hameed, Michael LeVatte, Ian Forsythe, and Reza~M. Salek.
\newblock Foodb: The food database.
\newblock \url{http://foodb.ca}, 2018.
\newblock Accessed: 2024-07-02.

\bibitem[Zhou et~al.(2020)Zhou, Cui, Hu, Zhang, Yang, Liu, and Sun]{zhou2020graph}
Jie Zhou, Guodong Cui, Shengding Hu, Zhiyuan Zhang, Cheng Yang, Zhiyuan Liu, and Maosong Sun.
\newblock Graph neural networks: A review of methods and applications.
\newblock \emph{AI Open}, 1:\penalty0 57--81, 2020.

\end{thebibliography}

\end{document}